\documentclass[%
 reprint,
 amsmath,amssymb,
 aps,
prl,
showkeys
]{revtex4-2}

\usepackage{graphicx}
\usepackage{dcolumn}
\usepackage{bm}
\usepackage[romanian,english]{babel}

\DeclareMathAlphabet\mathbfcal{OMS}{cmsy}{b}{n}
\newcommand{\prob}{p}
\newcommand{\vecprob}{\bm{p}}
\newcommand{\dens}{\rho}
\newcommand{\vecdens}{{\bm{\dens}}}
\newcommand{\traj}{\mathcal{X}}
\newcommand{\vectraj}{\mathbfcal{X}}
\newcommand{\probtraj}{\mathcal{P}}
\newcommand{\vecprobtraj}{{\mathbfcal{P}}}
\newcommand{\probtrans}{w}
\newcommand{\matprobtrans}{\mathbf{W}}
\newcommand{\vecstate}{\mathbf{x}}
\graphicspath{{./figs/},}

\begin{document}

\preprint{APS/123-QED}

\title{Measuring dynamical phase transitions in time series}

\author{Bulcs\'{u} S\'{a}ndor}
\email{bulcsu.sandor@ubbcluj.ro}
\affiliation{%
 Department of Physics, Babe\textcommabelow{s}-Bolyai University, Cluj-Napoca, Romania\\
}%
\affiliation{
 Transylvanian Institute of Neuroscience, Cluj-Napoca, Romania
}
\author{Andr\'as Rusu}
\affiliation{%
 Department of Physics, Babe\textcommabelow{s}-Bolyai University, Cluj-Napoca, Romania\\
}%
\affiliation{
 Transylvanian Institute of Neuroscience, Cluj-Napoca, Romania
}
\author{K\'aroly D\'enes}
\affiliation{%
 Department of Physics, Babe\textcommabelow{s}-Bolyai University, Cluj-Napoca, Romania\\
}%
\affiliation{
 Transylvanian Institute of Neuroscience, Cluj-Napoca, Romania
}
\author{M\'aria Ercsey-Ravasz}
\email{maria.ercsey@ubbcluj.ro}
\affiliation{%
 Department of Physics, Babe\textcommabelow{s}-Bolyai University, Cluj-Napoca, Romania\\
}%
\affiliation{
 Transylvanian Institute of Neuroscience, Cluj-Napoca, Romania
}
\author{Zsolt I. L\'az\'ar}
\email{zsolt.lazar@ubbcluj.ro}
\affiliation{%
 Department of Physics, Babe\textcommabelow{s}-Bolyai University, Cluj-Napoca, Romania\\
}%
\affiliation{
 Transylvanian Institute of Neuroscience, Cluj-Napoca, Romania
}




\date{\today}

\begin{abstract}
There is a growing interest in methods for detecting and interpreting changes in experimental time evolution data. Based on measured time series, the quantitative characterization of dynamical phase transitions at bifurcation points of the underlying chaotic systems is a notoriously difficult task. Building on prior theoretical studies that focus on the discontinuities at $q=1$ in the order-$q$ Rényi-entropy of the trajectory space, we measure the derivative of the spectrum.
We derive within the general context of Markov processes a computationally efficient closed-form expression for this measure. 
We investigate its properties through well-known dynamical systems exploring its scope and limitations. 
The proposed mathematical instrument can serve as a predictor of dynamical phase transitions in time series.

\end{abstract}

\maketitle


\paragraph*{Complex systems and time series---}
Recently, in the era of complexity science, the significance of efficient tools for characterizing quantitatively the temporal evolution of real-world complex systems has surged \cite{crutchfield2012,murphy2024}. 
In the absence of defining equations of motion the focus shifts towards analyzing the observed time series \cite{sakellariou2019,boaretto2021,bartlett2022}. 
Qualitative changes in the dynamics are important from the point of view of prediction and control. 
In biological time series, for example, accurately detecting epileptic seizures in EEG recordings and identifying cardiac fibrillation in ECG data are vital \cite{lehnertz2023,toker2020}.
Therefore, the study of critical transitions and the identification and location of tipping points are fundamental aspects of complex systems \cite{sandor2013,koeglmayr2024,morr2024}.
There are theoretical methods for describing the critical behavior of dynamical systems allowing for the definition and estimation of quantities with known analogs in thermodynamic formalism including entropy, partition function, free energy, and diffusion coefficient \cite{tel1988,beck1995thermodynamics}. 
These are mainly demonstrated in 1D systems covering both phase space \cite{schlogl1974,schlogl1988} and trajectory (history) space statistics \cite{szepfalusy1987,szepfalusy1989,csordas1989,fujisaka1983}. 
However, analytical relations can be exploited only in a limited number of cases.
For a numerical measurement of the entropy, a symbolic representation of time series is realized in one of two ways: spatial encoding (threshold-based phase space partitioning) or temporal encoding (ordinal patterns) \cite{rubido2018,sakellariou2019,zou2019,zanin2021}. 
As a result, the problem can be transferred to the context of symbolic dynamics, setting the stage for a computational and information-theoretical perspective \cite{crutchfield1990computation,boffetta2002}.

In the absence of equations defining the dynamics of the system, such as for Markov processes or experimental time series it is a challenge to estimate the relevant statistical quantities related to phase transitions. 
Here we briefly review the role of order-$q$ Rényi-entropy in characterizing dynamical properties as they manifest both in the phase space and trajectory space. 
Based on previous studies \cite{szepfalusy1987,szepfalusy1989,csordas1989,fujisaka1983} we propose to measure dynamical phase transitions via the derivative of the entropy at $q=1$ exhibiting singularities at tipping points. 
We derive a computationally efficient closed-form expression of this measure within the framework of a first-order Markov process model of the dynamics.
Here we present examples for one and two-dimensional maps, however, the method can be applied in higher dimensions, too. 
It can be generalized to continuous time dynamical systems and multivariate time series by adequate time discretization, for example by constructing the corresponding  Poincare map, as described in Refs.~\cite{sakellariou2019,sandor2021}.
Our approach bridges a current methodological gap by addressing both the theoretical and practical aspects of analyzing phase transitions in complex dynamical systems, with applicability spanning biological, socio-economic, geophysical, and climate-related time series.

\paragraph*{Generalized entropies.---}The generalized entropy (also known as order-$q$ R\'enyi entropy) for a discrete distribution $\vecprob=(p_1, p_2, \dots, p_N)$ is defined as
\begin{equation}
H_q(\vecprob) = \frac{1}{1-q}\ln\sum_{i=1}^N\prob_i^q
\label{eq:renyi_general}
\end{equation}
where $H_{q}\geq H_{q'}$ whenever $q'>q$ \cite{beck1995thermodynamics}. In the $q=1$ limit, we get the mean 
\begin{equation}
H_1(\vecprob) = -\langle\ln\vecprob\rangle_{\vecprob}=-\sum_{i=1}^N p_i\ln p_i\,,
\label{eq:shannon_general}
\end{equation}
which is the Shannon entropy for the same distribution. 
The same limit of the derivative of the Rényi-entropy, with respect to $q$, gives~\cite{Buryak2021}
\begin{equation}
-2H_1'(\vecprob) = \Delta^2_{\vecprob}\ln(\vecprob) = \langle\ln^2\vecprob\rangle_{\vecprob} -\langle\ln\vecprob\rangle^2_{\vecprob}\ ,
\label{eq:renyi_general_derivative}
\end{equation}
hence, it is proportional to the bit-number variance of the distribution, $\Delta^2_{\vecprob}\ln(\vecprob)$ \cite{schlogl1974,beck1995thermodynamics}. 
Depending on the context, the same quantity is also referred to as diffusion coefficient \cite{fujisaka1983} or specific heat \cite{schlogl1988} depending on whether $\ln\vecprob$ is associated with displacement during a random walk or energy of a thermodynamic system. 

\paragraph*{Dynamical systems.---}
For discrete-time dynamical systems, we need an appropriate partitioning of the phase space by defining a set of $n$ nonoverlapping cells that cover the attractor entirely. 
In the limit of infinitesimal cell sizes, Eq.~(\ref{eq:renyi_general}) can be used to estimate the resolution-independent spectrum of R\'enyi dimensions.
The Shannon entropy and the bit-number variance can be defined using the coarse-grained approximation $\vecdens^*=(\dens_1^*, \dens_2^*, \dots, \dens_n^*)$ of the corresponding natural distribution on the attractor,
\begin{equation}
C_1 =  H_1(\vecdens^*)\,,\qquad
C_2 = -2 H'_1(\vecdens^*)
\label{eq:bit_numbers}
\end{equation}
where the latter exhibits singularities at bifurcation points \cite{schlogl1974,schlogl1988}. While the natural distribution can be estimated numerically its analytical expression is only known for some parameters of a limited number of maps. 
For time series, in the absence of the map, the quantities $C_1$ and $C_2$ remain invariant under the reshuffling of the data points, so it is not suitable for measuring the more relevant trajectory-level information \cite{boffetta2002}. 
To that end a trajectory of the system can be defined as a string of symbols, viz.\ coarse-grained states $\traj(t)=\{s_1, s_2, \dots, s_t\}$ if $\vecstate(\tau=1)\in$ cell $s_1$, $\vecstate(\tau=2)\in$ cell $s_2$, $\dots$, $\vecstate(\tau=t)\in$ cell $s_t$ where $\vecstate(\tau)$ denotes the state of the dynamical system at time step $\tau$. 
Let us denote by $\probtraj(\traj(t))$ the joint probability of the trajectory $\traj(t)$ of $t$ consecutive steps. 
The corresponding dynamic entropy in the probability space $\vecprobtraj(t)$ of all possible trajectories $\vectraj(t)$ is defined as \cite{grassberger1983measuring,grassberger1983characterization,grassberger1984dimensions}:
\begin{equation}
H_q(\vecprobtraj(t)) = \frac{1}{1-q}\ln\sum_{\vectraj(t)}\left[\probtraj(\traj(t))\right]^q 
\end{equation}
While the above expression usually diverges with the length of the trajectories $t$, the dynamical R\'enyi entropy rate,
\begin{equation}
K_q(\vecprobtraj) = \lim_{t\rightarrow\infty}\frac{1}{t} H_q(\vecprobtraj(t))\,,
\label{eq:renyi_traj}
\end{equation}
characterizes the scaling behavior in the limit of infinite time. The above quantity generally depends on the partitioning. Using the generating partition the supremum over partitions is reached.  As such a partitioning is generally not known, one may choose instead a grid of cells of equal size. A sufficiently high resolution is also expected to provide a good estimation of the supremum~\cite{cohen1985}. Alternatively, one may also use numerical or machine-learning algorithms for estimating optimal partitions from observed data (see e.g.~\cite{kennel2003,hirata2004,murphy2024}).

The Rényi-entropy at $q=1$, denoted here by $S$, is of special interest and it is called the Kolmogorov-Sinai (KS) entropy or metric entropy~\cite{boffetta2002}:
\begin{equation}
S = K_1(\vecprobtraj)\,,
\qquad
\Lambda = -{2}K_1'(\vecprobtraj)\,,
\label{eq:renyi_derivative}
\end{equation}
as it measures the entropy production rate due to the chaotic nature of the dynamics. Based on \cite{szepfalusy1987,kaufmann1989} there exists a dynamical phase transition at borderline cases of chaos, hence the derivative of the Rényi-entropy with respect to $q$ at $q=1$ (interpreted as diffusion coefficient), quantified by $\Lambda$ in Eq.~(\ref{eq:renyi_derivative}), diverges at the critical point. 

\paragraph*{Markov chains.---}
Discrete-time autonomous dynamical systems with a continuous state space can be approximated by time-homogeneous discrete-time Markov chains with finite state space. This model requires two elements. Firstly, the states are discretized by choosing a suitable partitioning as described above. 
Secondly, the transition probabilities between states $i$ and $j$ are defined as a conditional probability $\probtrans_{ij}=p(s_{t+1}=j|s_{t}=i)$, where $i,j=1,\dots,N\leqslant n$, where $n$ is the number of cells and $N$ is the number of symbols present in the encoded time series. 
Given such a Markov chain with the corresponding transition matrix, $(\matprobtrans)_{ij}=\probtrans_{ij}$, we call its network representation a state-transition network (STN) \cite{sandor2021}. 
This approximation becomes exact if the coarse-graining corresponds to Markov partitioning, one that is only known exactly for a handful of systems and can be approximated or optimized numerically in some other cases as well~\cite{rubido2018}. 
Additionally, the first-order Markov chain description of the dynamics can be improved by considering higher-order states corresponding to symbolic sequences $s_{t-m+1},\dots,s_t$ of length $m$ which leads to an $m$-th order Markov process, $p(s_{t-m+2},\dots,s_{t+1}|s_{t-m+1},\dots,s_t) =
p(s_{t+1}|s_{t-m+1},\dots,s_t)$
in terms of the original symbolic states.
The number of symbols encoding $m$-th order states with non-zero probability during the dynamics is typically $N\ll n^m$.
The stationary distribution of the Markov chain, $\vecdens$, is the eigenvector of the transition probability matrix $\matprobtrans$ associated with the unit eigenvalue (see Section B in the Supplemental Material \cite{supp}). 
In the special case of first-order states, $m=1$, it coincides with the coarse-grained natural distribution on the attractor,~$\vecdens=\vecdens^*$. 

For estimating the dynamical entropies $K_q$, we define the trajectory $\traj(t)$ as a sequence of states/nodes visited during a random walk of $t$ steps on the network, i.e.\ a realization of the Markov chain. The probability of such a trajectory can be given as
\begin{equation}
\probtraj(\traj(t)) = \dens_{s_1}\probtraj(\traj(t)|s_1)\,,
\quad
\probtraj(\traj(t)|s_1) = \prod_{\tau=1}^{t-1} (\probtrans_{s_\tau s_{\tau+1}}) \,.
\label{eq:trajectory_probability}
\end{equation}
Using the Markov-chain description, the full spectrum of the truncated R\'enyi-entropy \cite{szepfalusy1986entropy,szepfalusy1989,kaufmann1989} can be given as (see e.g.\ \cite{wu2023} and Section C in the Supplemental Material~\cite{supp}).
\begin{equation}
\tilde{K}_q = {\ln\alpha_\text{max}(q)}/\left({1-q}\right)
\label{eq:renyi_entropy_estimation}
\end{equation}
where $\alpha_\text{max}(q)$ is the largest eigenvalue of the $(\matprobtrans_q)_{ij}=\probtrans_{ij}^q$ matrix (element-wise $q$ power). 
The truncated entropy, $\tilde{K}_q$, is expected to exceed the exact value, $K_q$, for reasons that are straightforward to show in the $q=1$ case from the properties of conditional entropies (see also Ref.~\cite{szepfalusy1986entropy}). Section G in the Supplemental Material \cite{supp} demonstrates Eq.~(\ref{eq:renyi_entropy_estimation}) for two prototypical one-dimensional systems: the piece-wise linear asymmetric tent map~\cite{beck1995thermodynamics} for which the analytically known R\'enyi-entropy spectrum is reproduced numerically, and in the case of the critical map~\cite{szepfalusy1987} exhibiting intermittent dynamics the discontinuity at $q=1$ is confirmed together with the exact value of the KS entropy.

For memory-less dynamics the truncated estimation of the entropy rate, $S$, and diffusion coefficient, $\Lambda$, from Eq.~(\ref{eq:renyi_derivative}) turn into the bit-number statistics, $C_1$ and $C_2$, from Eq.~(\ref{eq:bit_numbers}) (see Section C in the Supplemental Material \cite{supp}). 

\begin{figure}[ht!]
\includegraphics[width=0.93\columnwidth,keepaspectratio]{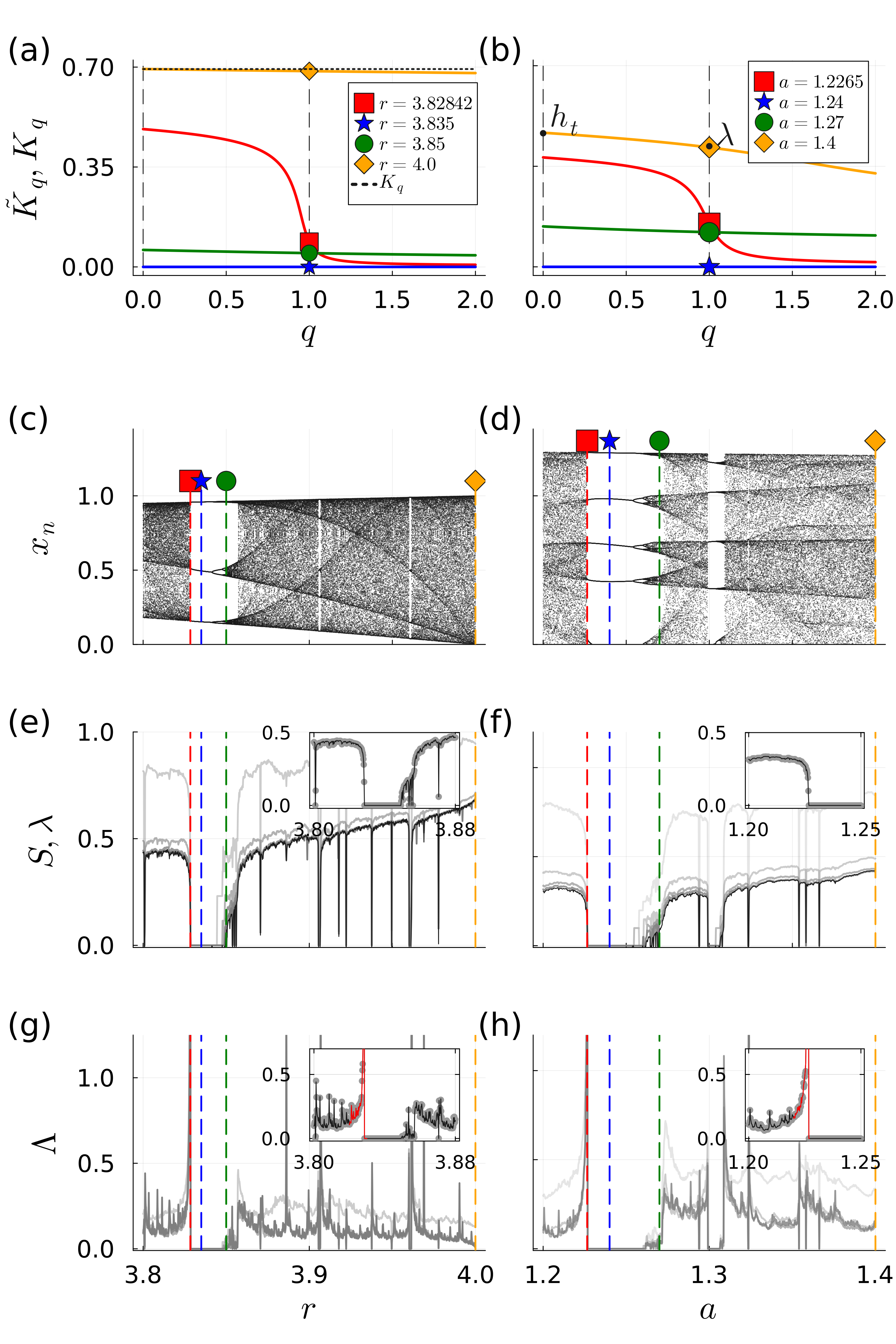}
\vspace{-0.5cm}
\caption{\label{fig:measures}
Matching the classical picture of the logistic map (left column) and the Henon map (right column) to its Markov-chain-based model. For the logistic map the [0,1] interval, while for the Henon map the phase plane $[-2, 2]\times[-2, 2]$ is partitioned into $n=2^5$ and $n = 2^5 \times 2^5$ equally sized cells, respectively.
(a,b) The spectrum of the truncated R\'enyi entropy $\tilde{K}_q$ computed by using Eq.~(\ref{eq:renyi_entropy_estimation}) and states of order $m=12$ for four different parameter values (given in the legends). 
The location of markers corresponds to the KS entropy $S=\tilde{K}_1$ given by Eq.~(\ref{eq:KS_entropy}).
The parameter values $r$ and $a$ used as examples in panels (a) and (b), respectively, are indicated by the corresponding markers and the vertical dashed lines in the panels below.
 (a) For the logistic map, the thin dashed line denotes the theoretical R\'enyi entropy, $K_q=\ln 2$, for $r=4$ \cite{beck1995thermodynamics}. 
(b) For the Henon map the topological entropy $h_\mathrm{t}\approx 0.465\approx \tilde{K}_0$ and the maximal Lyapunov exponent $\lambda\approx 0.42\approx S$ are shown for $a=1.4$ for comparison~\cite{dAlessandro_1990}. 
(c,d) The bifurcation diagrams of the above systems (see Sec.~G in the Suppl.\ Mat.\ \cite{supp} for a broader range of the control parameters).  
(e,f) The KS entropy, $S(m)$, and (g,h) the Lyapunov measure, $\Lambda(m)$, from Eqs.~(\ref{eq:KS_entropy}) and (\ref{eq:lyapunov_analytic}), respectively. Lines in increasingly darker shades of gray correspond to Markov chains using states of order $m = 1,2,4,12$, respectively.
The analytical results (gray lines) given in Eq.~(\ref{eq:lyapunov_analytic}) are compared to the 
statistics of $10^3$ simulated random walks (gray circular markers) on the corresponding STNs in the insets for order $m=12$. 
In panels (e,f) the largest Lyapunov exponent, $\lambda$, is also shown by the thin black line for comparison. 
Note that $S(12)\approx\lambda$ for $\lambda>0$.
}
\end{figure}

In the case of the logistic map, $x_{\tau+1}=rx_\tau(1-x_\tau)\,,$
the truncated entropy exhibits a phase transition-like discontinuity at $q=1$ only in the vicinity of critical values of the parameter $r$ corresponding to bifurcation points (e.g.\ $r=3.82842$, compare the lines denoted by the square-shaped marker in panels (a) and (c) of Fig.~\ref{fig:measures}). 
Around these points, the evolution of the system presents properties characteristic of phase transitions in physical systems such as critical slowing down. 
In the case of limit cycles, the full entropy spectrum is identically zero, for partially predictable chaos~\cite{wernecke2017test}, however, it is characterized by relatively low, quasi-constant values (see the star- and circle-shaped markers in panel (a) of Fig~\ref{fig:measures}, respectively).
Finally, for $r=4$ viz for fully developed chaos, all R\'enyi entropies are $K_q=\ln 2$ (see the dashed line in Fig~\ref{fig:measures}(a) and its numerical estimate denoted by the yellow line with the diamond-shaped marker) \cite{beck1995thermodynamics}, which is exactly reproduced by its truncated version in Eq.~(\ref{eq:renyi_entropy_estimation}) when using the Markov partitions [0, 0.5) and [0.5, 1) (not shown here).

The same can be done if the map is higher-dimensional. Here, we consider the well-known two-dimensional Henon map \cite{Alligood1996}, $x_{\tau+1} = 1- ax_\tau^2+y_\tau \,,\, y_{\tau+1} = b x_\tau \,,$ with $b=0.3$ and control parameter $a$.
A similar phase transition can be observed in the vicinity of the critical values of the control parameter (e.\ g.\ $a\approx 1.2265$, compare panels (b) and (d) of Fig.~\ref{fig:measures}), see also in \cite{sandor2021}. 
Note that for $a=1.4$ the accepted value of the topological entropy $h_\mathrm{t}\approx 0.465$ is reproduced by $\tilde{K}_0\approx h_\mathrm{t}$ \cite{dAlessandro_1990,sakellariou2021}. Furthermore, the KS entropy is equal to the maximal Lyapunov exponent, $\tilde{K}_1\approx \lambda\approx 0.42$, as expected by Pesin's relation~\cite{boffetta2002}.

The critical behaviors illustrated above can be described by the quantity $\Lambda$ in Eq.~(\ref{eq:renyi_derivative}) corresponding to specific heat in thermodynamics and associated with the bit number variance of the trajectory probabilities, see Eq.~(\ref{eq:renyi_general_derivative}). 
As it is shown in the next section, this bit number statistics has an intuitive interpretation and allows for a complete analytical description within the framework of Markov chains. 

\paragraph*{Random-walks statistics.---}
The two measures from Eq.~(\ref{eq:renyi_derivative}) can be expressed as the mean and variance of  
\begin{equation}
\mathcal{L}(t) \equiv -\ln\mathcal{P}(\mathcal{X}(t)|s_1)  
\end{equation}
over all trajectories $\traj(t)$ and initial states $s_1$ (see Section D in Supplemental Material \cite{supp}).
They can be estimated within the framework of first-order Markovian processes discussed above by using the truncated entropy $\tilde{K}_q$. 
According to Eq.~(\ref{eq:trajectory_probability}) the quantity $\mathcal{L}(t)$ can be interpreted as the total length of a random walk trajectory, $\traj(t)$, made up of steps of length $l_\tau=-\ln\probtrans_{s_\tau s_{\tau+1}}$ with $\mathcal{L}(t)=\sum_{\tau=1}^t l_\tau$.

Due to the Markov chain central limit theorem~\cite{jones2004}, the distribution of total path lengths, $\mathbfcal{L}(t)$, over all possible realizations of random walks, $\vectraj(t)$, approaches a normal distribution in the limit of $t\rightarrow\infty$, which can be fully characterized by its mean, $\langle \mathbfcal{L}(t)\rangle_\vecprobtraj$, and variance, $\Delta_\vecprobtraj^2 \mathbfcal{L}(t)$ (see Section G in the Supplemental Material \cite{supp}). 
These two quantities describing the statistics of random-walk path lengths turn out to be closely related to the phase-transition measures proposed in Eq.~(\ref{eq:renyi_derivative}) as discussed below.

The truncated estimation of the entropy rate from Eq.~(\ref{eq:renyi_derivative}) can be given as the time-normalized mean path length, $\langle \mathbfcal{L}(t)\rangle_\vecprobtraj/t$ (see Sections D and F in Supplemental Material \cite{supp}),
which reduces to the Kolmogorov-Sinai entropy rate of the underlying Markov process~\cite{sandor2021}:
\begin{equation}
S = \langle l\rangle
 = -\sum_{i,j=1}^N \dens_i\probtrans_{ij} \ln \probtrans_{ij} = \vecdens^\top \mathbf{L}_1 \mathbf{1}\,,
\label{eq:KS_entropy}
\end{equation}
where $\langle \cdot \rangle$ denotes the average over the ensemble of random walkers, $\vecdens^\top$ and $\mathbf{1} = (1, 1, \dots, 1)$ are the left and right eigenvectors of the stochastic matrix $\matprobtrans$ with corresponding eigenvalue $\alpha_1 = 1$, and $(\mathbf{L}_1)_{ij}=-\probtrans_{ij}\ln \probtrans_{ij}$ is the probability-weighted edge-length matrix.
In the context of STNs, this entropy is equivalent to the average of the entropies associated with individual nodes (see Section E in Supplemental Material \cite{supp}).

Similarly, the time-normalized variance, $\Delta_\vecprobtraj^2 \mathbfcal{L}(t)/t$, of random-walk path lengths on the same STN is given by the Lyapunov measure (see Sections D and F in Supplemental Material \cite{supp}),
\begin{equation}
\Lambda = 
\sigma_l^2 +2\sum_{\tau=1}^\infty \Bigl( \langle l_1 l_{1+\tau} \rangle - \langle l_1\rangle \langle l_{1+\tau}\rangle\Bigr)\,,
\end{equation}
where the first term denotes the variance of the individual edge lengths,
\begin{equation}
\sigma_l^2 = \langle l^2 \rangle - \langle l\rangle^2 
\,,
\qquad
\langle l^2\rangle = \sum_{i,j=1}^N \dens_i\probtrans_{ij} \ln^2 \probtrans_{ij}\,,
\end{equation}
sampled according to the Markov process, while the second term is the sum of auto-covariances for all possible time lags $\tau$, with $l_1$ being the edge length in the first step~\cite{brooks2011handbook}.
Using the transition probability matrix $\matprobtrans$, the Lyapunov measure can be expressed in finite form (see Section F in Supplemental Material \cite{supp}) as
\begin{equation}
\Lambda =  \vecdens^\top \mathbf{{L}}_2 \mathbf{1} - (\vecdens^\top \mathbf{L}_1 \mathbf{1})^2 + 2\vecdens^\top \mathbf{L}_1 \mathbf{B} \mathbf{L}_1 \mathbf{1}\,,
\label{eq:lyapunov_analytic}
\end{equation}
where $(\mathbf{{L}}_2)_{ij}=\probtrans_{ij}\ln^2 \probtrans_{ij}$, $\mathbf{B} \equiv (\mathbf{I}-\matprobtrans+\mathbf{1}\vecdens^\top)^{-1} - \mathbf{1}\vecdens^\top$.

The equivalence of the closed-form expressions of the two measures in Eqs.~(\ref{eq:KS_entropy}), (\ref{eq:lyapunov_analytic}) and the corresponding path length statistics obtained from the direct simulation of the underlying random walk is demonstrated for the logistic and the Henon maps in the insets of panels (e), (g) and (f), (h),  respectively, of Fig.~\ref{fig:measures}. 
As expected based on Pesin's relation the Kolmogorov-Sinai entropy (gray lines in Fig.~\ref{fig:measures}(e), (f)) matches the Lyapunov exponent, $\lambda$ (black line), computed for all control parameter values using Benettin's method \cite{benettin1976,datseris2018}. 
The two get closer as the order, $m$, of the states increases (darker shades of gray), $S(m)\rightarrow\lambda$.
However, the Lyapunov measure tends to diverge in the neighborhood of certain bifurcation points irrespective of the details of the partitioning and the used order, $m$, for symbolic encoding (see the red peaks in the insets of panels Fig.~\ref{fig:measures}(g), (h). 
On the other hand, the width and height of the peaks can be controlled via the applied resolution, $n$,  (not shown here) and order, $m$.
Systems undergoing parameter drift are expected to produce similar divergences during their quasi-stationary evolution prior to a dynamical phase transition. 
Hence an abrupt increase in the Lyapunov measure, for example preceding the collapse of chaos to periodic behavior, can serve as a precursor of such tipping points.

The proposed method for measuring dynamical phase transitions expressed by Eqs.~(\ref{eq:KS_entropy}) and (\ref{eq:lyapunov_analytic}) is based on common matrix algebra at a computational cost of $\mathcal{O}(N^3)$, and storage cost of $\mathcal{O}(N^2)$ where $N$ is the number of symbols. 
For large $N$ (high phase space resolution, $n$, or high-order representations, $m$) the transition matrix, $\matprobtrans$, is typically sparse allowing for low complexity iterative numerical methods (see Sections H and I for the iterative method and code availability, respectively, of the Supplemental Material \cite{supp}).

\paragraph*{Conclusions.---}
The proposed Lyapunov measure based on the first-order Markov process model of dynamical systems can detect dynamical phase transitions manifesting as discontinuities in the Rényi-entropy (e.g.\ at weak intermittency). 
For memory-less dynamics, the KS entropy and the Lyapunov measure reduce to the bit-number statistics $C_1$ and $C_2$ from Eq.~(\ref{eq:bit_numbers})  which rely solely on state distribution. 
However, for general dynamics, the former two measures calculated from trajectories convey additional information on the underlying behavior of the system telling apart meaningful dynamics from random noise with the same statistics~\cite{rosso2007,boaretto2021}.

The formalism developed here allows for any symbolic representation of the time series including ordinal partition networks \cite{zou2019}. 
The accuracy of the numerical estimates of entropy-related quantities is solely dependent on the intrinsic limitations of the employed representation, compare, for example, the results on topological entropy using the ordinal patterns representations of the Henon-map~\cite{sakellariou2021}.
The proposed method can be employed for characterizing dynamical systems and time series even at high-resolution coarse graining and high-order symbolic representation allowing cost-effective numerical estimation of the two statistical measures. 
Our transition matrix-based procedure can be applied in higher dimensions and it can be generalized to continuous time dynamical systems and multivariate time series. 
 
In light of the recent surge in experimental data, there is a growing need for tools that can extract useful information. We have outlined the mathematical foundations of a method that can serve as an effective instrument in this regard, enabling the measurement of dynamical phase transitions in time series and facilitating the detection of early warning signals of critical transitions in chaotic dynamics with immediate applications in areas such as finances, climate research and  neuroscience.

\medskip
\begin{acknowledgments}
This work was supported by the grant of the Romanian Ministry of Research, Innovation and Digitization, CNCS - UEFISCDI, project number PN-III-P4-PCE-2021-0408 (BS, KD, MER, ZL), PN-III-P4-ID-PCE-2020-0647 (BS, ZL), ERANET-FLAG-ERA-ModelDXConsciousness (BS, MER, ZL), ERANET-NEURON-2-RESIST-D (MER, BS, AR, ZL), ERANET-NEURON-2-IBRAA (MER), ERANET-FLAG-ERA-JTC2023-MONAD (MER), ERANET-NEURON-UnscrAMBLY (MER) within PNCDI III, and SRG-UBB 32993/23.06.2023 (BS) within UBB Starting Research Grants of the Babe\textcommabelow{s}-Bolyai University, and by the Collegium Talentum Programme of Hungary (AR). The authors also acknowledge the useful discussion with Tam\'as T\'el.
\end{acknowledgments}



\providecommand{\noopsort}[1]{}\providecommand{\singleletter}[1]{#1}%

\end{document}


\preprint{APS/123-QED}

\title{SUPPLEMENTAL MATERIAL\\Measuring dynamical phase transitions in time series}

\author{Bulcs\'{u} S\'{a}ndor}
\email{bulcsu.sandor@ubbcluj.ro}
\author{Andr\'as Rusu}
\author{K\'aroly D\'enes}
\author{M\'aria Ercsey-Ravasz}
\email{maria.ercsey@ubbcluj.ro}
\author{Zsolt I. L\'az\'ar}
\email{zsolt.lazar@ubbcluj.ro}
\affiliation{%
 Department of Physics, Babe\textcommabelow{s}-Bolyai University, Cluj-Napoca, Romania\\
 Transylvanian Institute of Neuroscience, Cluj-Napoca, Romania
}%




\date{\today}

\begin{abstract}
There is a growing interest in methods for detecting and interpreting changes in experimental time evolution data. Based on measured time series, the quantitative characterization of dynamical phase transitions at bifurcation points of the underlying chaotic systems is a notoriously difficult task. Building on prior theoretical studies that focus on the discontinuities at $q=1$ in the order-$q$ Rényi-entropy of the trajectory space, we measure the derivative of the spectrum.
We derive within the general context of Markov processes a computationally efficient closed-form expression for this measure. 
We investigate its properties through well-known dynamical systems exploring its scope and limitations. 
The proposed mathematical instrument can serve as a predictor of dynamical phase transitions in time series.

\end{abstract}

\maketitle


\subsection{Entropies}
Here we summarize the entropy-related definitions from the main paper. 
The  order-$q$ R\'enyi entropy for a discrete distribution $\vecprob=(p_1, p_2, \dots, p_N)$ is defined as
\begin{equation}
H_q(\vecprob) = \frac{1}{1-q}\ln\sum_{i=1}^N\prob_i^q\,,
\label{eq:renyi_general}
\end{equation}
where $H_{q}\geq H_{q'}$ whenever $q'>q$ 
. 
The Shannon entropy of the same distribution can be obtained in the $q=1$ limit, as the mean 
\begin{equation}
H_1(\vecprob) = -\langle\ln\vecprob\rangle_{\vecprob}=-\sum_{i=1}^N p_i\ln p_i\,.
\label{eq:shannon_general}
\end{equation}
The same limit of the derivative of the Rényi-entropy, with respect to $q$, gives the variance
%
\begin{equation}
H_1'(\vecprob)=-\frac{1}{2}\Delta^2_{\vecprob}\ln(\vecprob)=-\frac{1}{2}\left(\langle\ln^2\vecprob\rangle_{\vecprob} -\langle\ln\vecprob\rangle^2_{\vecprob}\right)\,.
\label{eq:renyi_general_derivative}
\end{equation}
For a dynamical system with coarse-grained natural distribution $\vecdens^*$ on the attractor we denote with
\begin{equation}
C_1 =  H_1(\vecdens^*)\,,\qquad
C_2 = -2 H'_1(\vecdens^*)\,.
\label{eq:bit_numbers}
\end{equation}
The dynamical R\'enyi-entropy is defined as the time-normalized entropy of the trajectory probabilities $\vecprobtraj$:
\begin{equation}
K_q(\vecprobtraj) = \lim_{t\rightarrow\infty}\frac{1}{t} H_q(\vecprobtraj(t))\,.
\label{eq:renyi_traj}
\end{equation}
%
%
The $q=1$ value is of special interest, and it is called the Kolmogorov-Sinai entropy or metric entropy~\cite{boffetta2002}:
\begin{equation}
S = K_1(\vecprobtraj)\,,
\qquad
\Lambda = -{2}K_1'(\vecprobtraj)\,.
\label{eq:renyi_derivative}
\end{equation}
%


\subsection{Markov chain formalism}
\label{A:markov_chain}
Left and right eigenvectors, $\vecdens_\nu^\top$ and $\mathbf{v}_\nu$, respectively, of the stochastic matrix $\matprobtrans=(\probtrans_{ij})$, with $i,j=1,\dots,N$, are defined as:
\begin{equation}
    \matprobtrans^\top \vecdens_\nu=\alpha_\nu\cdot\vecdens_\nu\,,\qquad\qquad
    \matprobtrans \mathbf{v}_\nu=\alpha_\nu\cdot\mathbf{v}_\nu\,,
\end{equation}
with $\alpha_\nu$ eigenvalues, where $\alpha_1=1$, and $\vecdens_1$ is the stationary distribution of the corresponding Markov chain, for which the normalization is given by $\vecdens_1^\top\mathbf{v}_1=1$ with $\mathbf{v}_1= (1,1,\dots,1)$.
Note that as symbols with zero probability play no role in this representation, the stationary distribution of the Markov chain $\vecdens_1$, for states of order $m=1$, provides an approximation of the natural distribution on the attractor $\vecdens^*$ when omitting cells with zero probability, $\vecdens_1=\vecdens^*$.

A useful identity concerning the powers of the transition matrix is
\begin{equation}
    \left(\matprobtrans-\alpha_\nu \mathbf{v}_\nu\vecdens_\nu^\top\right)^k=\matprobtrans^k-\alpha_\nu^k\mathbf{v}_\nu\vecdens_\nu^\top\,,
    \qquad\quad
    k \geq 1\,.
    \label{eq:power_identity}
\end{equation}
In the following we denote $\vecdens=\vecdens_1$ and $\mathbf{1} =\mathbf{v}_1$ for simplicity.

\subsection{R\'enyi entropy spectrum}
\label{A:renyi_spectrum}
The full spectrum of the truncated R\'enyi-entropies can be given as
\begin{eqnarray}
\tilde{K}_q = \frac{1}{1-q}\lim_{k\rightarrow\infty} \frac{1}{k} \ln \left(\sum_{i,j} \dens_i^q (\matprobtrans_q^k)_{ij}1_j\right)= 
\nonumber\\
\frac{1}{1-q}\lim_{n\rightarrow\infty}\ln \left(\frac{\sum_{i,j} \dens_i^q (\matprobtrans_q^{k+1})_{ij}1_j}{\sum_{i,j} \dens_i^q (\matprobtrans_q^{k})_{ij}1_j}\right) =
\nonumber\\ 
\frac{\ln\alpha_\text{max}(q)}{1-q}\,,
\label{eq:renyi_entropy_estimation_derivation}    
\end{eqnarray}
where $\alpha_\text{max}(q)$ is the largest eigenvalue of the $(\matprobtrans_q)_{ij}=\probtrans_{ij}^q$ matrix (element-wise $q$ power). 
According to the Perron-Frobenius theorem for irreducible non-negative matrices the spectral radius $\alpha_\text{max}$ is a positive real number. 

For random systems (without memory), viz. for $\probtrans_{ij}=1_i\dens_j$, where $1_i$ denotes the $i$-th component of a vector of all-ones, the above expression reduces to the R\'enyi entropy of the stationary distribution,
\begin{equation}
\tilde{K}_q = \frac{1}{1-q}\ln \sum_{i=1}^N \dens_i^q = H_q(\vecdens)\,.
\label{eq:}
\end{equation}
Hence, for memory-less dynamics and first-order states, the truncated estimation of the entropy rate, $S$, and diffusion coefficient, $\Lambda$, from Eq.~(\ref{eq:renyi_derivative}) turn into the bit-number statistics, $C_1$ and $C_2$, from Eq.~(\ref{eq:bit_numbers}). 


\subsection{Path length statistics and entropies}\label{A:path_length_statistics}
Using the notations  $\vecprobtraj(t)\equiv \dens_{s_1}\vecprobtraj(\vectraj(t)|s_1)$ for the set of probabilities of all trajectories of $t$ steps,
 and $\mathbfcal{L}(t)\equiv-\ln\vecprobtraj(\vectraj(t)|s_1)$ for the length of trajectories with given initial state, Eqs.~(\ref{eq:renyi_general}) and (\ref{eq:renyi_traj}) can be combined into
\begin{align}
{K}_1(\vecprobtraj) &=-\lim_{t\rightarrow\infty}\frac{1}{t}\langle\ln\vecprobtraj(t)\rangle_\vecprobtraj=\nonumber\\
&=-\lim_{t\rightarrow\infty}\frac{1}{t}\left({\langle\ln\vecdens\rangle_{\vecdens}} - \langle\mathbfcal{L}(t)\rangle_\vecprobtraj\right)=
\nonumber\\
&=\lim_{t\rightarrow\infty}\frac{1}{t}\langle\mathbfcal{L}(t)\rangle_\vecprobtraj\,,
\label{eqapp:K1}
\end{align}
where averaging is done over all states~$s_i$, realizations of trajectories $\traj(t)$ of length $t$ and initial states $s_1$, respectively. 

Similarly, Eqs.~(\ref{eq:renyi_general_derivative}) and (\ref{eq:renyi_traj}) yield
\begin{equation}
{K}_1'(\vecprobtraj)= -\lim_{t\rightarrow\infty}\frac{1}{2t}\Delta_\vecprobtraj^2\ln(\vecprobtraj(t))\,.
\label{eqapp:K1der}
\end{equation}
Since generally $\Delta^2(x + y)= \Delta^2 x + \Delta^2 y + 2 \text{Cov}(x, y)$ for random variables $x$ and $y$, and the covariance between $\dens_i$ and $\probtraj(\traj(t)|s_1)$ goes to zero as the number of steps, $t$,  goes to infinity, $\mathrm{Cov}(\dens_i,\probtraj(\traj(t)|s_1))\rightarrow 0$, we get
\begin{align}
{K}_1'(\vecprobtraj) &= -\lim_{t\rightarrow\infty}\frac{1}{2t}\left(\Delta_\vecdens^2\ln\vecdens + \Delta_\vecprobtraj^2 \mathbfcal{L}(t)\right)=\nonumber\\
&=-\lim_{t\rightarrow\infty}\frac{1}{2t}\Delta_\vecprobtraj^2 \mathbfcal{L}(t)\,.
\end{align}

\subsection{Interpretation of the KS entropy}
\label{A:KS_interpretation}
The entropy rate determines the mean of the normal distribution, see Eq.~(\ref{eqapp:K1}), and can be interpreted as the amount of uncertainty removed by evolving the Markov chain with one step:
\begin{equation}
S = \sum_{i}^N \dens_i S_i
\,,
\qquad
S_i = -\sum_{j}^N \probtrans_{ij}\ln \probtrans_{ij}\,,
\label{eq:node_entropy}
\end{equation}
averaged over all state entropies $S_i$.
The KS entropy can be decomposed into the difference between the transition's entropy and the entropy of the stationary distribution of states:
\begin{equation}
S = S_\mathrm{tr} - C_1
\end{equation}
where 
\begin{equation}
S_\mathrm{tr} = -\sum_{i,j}^N \dens_i\probtrans_{ij} \ln \left( \dens_i \probtrans_{ij}\right)\,,
\quad
C_1 = -\sum_{i}^N \dens_i \ln \dens_i
\end{equation}
are respectively the transition entropy and state entropy, compare Eq.~(\ref{eq:bit_numbers}).
Note that using transition probabilities and $q=1$ we get a proper extensive (in time) statistical physics.
The mean bit number $C_1$ has, however, no straightforward interpretation as an extensive quantity.





\subsection{Lyapunov measure as ensemble averages}
\label{A:ensemble_averages}
In order to compute the Lyapunov measure $\Lambda$, first we need to estimate the auto-covariances,
\begin{equation}
\mathrm{Cov}(l_1,l_{1+\tau}) = \langle l_1 l_{1+\tau} \rangle - \langle l_1\rangle \langle l_{1+\tau}\rangle
\,,
\label{eq:cov}
\end{equation}
for all possible time-lags $\tau$, where $l_1$ is the edge length in the first step.
Due to the stationarity of the process, considering an ensemble of random walkers with initial distribution $\vecdens$, the ensemble-averaged step length is constant along the $\tau=1,\dots,t$ steps of the random walks, $\langle l\rangle = \langle l_{\tau}\rangle$, and it can be expressed as:
\begin{eqnarray}
\langle l\rangle = 
-\sum_{i=1}^{N} \dens_i \sum_{j=1}^{N} \probtrans_{ij} \ln\probtrans_{ij} = 
\vecdens^\top \mathbf{L}_1 \mathbf{1}\,,
\label{eq:l_av}
\end{eqnarray}
where $(\mathbf{L}_1)_{ij}=-\probtrans_{ij}\ln\probtrans_{ij}$ is the probability-weighted edge-length matrix. Note also that for $t\gg 1$ the ensemble-averaged edge lengths during the random walks do not depend on the starting positions of the walks.

Now, considering time-lags $\tau\in\{1,2,\dots\}$ we calculate the $\tau$-th order covariance of step-lengths $l$,
\begin{align}
\langle l_{1} l_{{1+\tau}} \rangle =& 
\nonumber\\
=&\sum\limits_{s_1=1}^{N} \dens_{s_1} \sum\limits_{s_2=1}^{N} \probtrans_{s_1s_2}(-\ln\probtrans_{s_1s_2}) \sum\limits_{s_3=1}^{N} \probtrans_{s_2s_3} \dots \nonumber\\
&...\sum\limits_{s_{\tau+1}=1}^{N} \probtrans_{s_\tau s_{\tau+1}} (-\ln\probtrans_{s_\tau s_{\tau+1}}) = 
\nonumber\\
&=\vecdens^\top \mathbf{L}_1 \matprobtrans^{\tau-1} \mathbf{L}_1 \mathbf{1}\,.\nonumber
\end{align}

Using identity (\ref{eq:power_identity}) with $\alpha_1=1$, for the infinite sum of covariances we need to calculate
\begin{eqnarray}
\mathbf{B} = \sum_{\tau=1}^\infty (\matprobtrans^{\tau-1}-\matprobtrans^{\infty}) = 
\sum_{k=0}^\infty (\matprobtrans^{k}-\mathbf{1}\vecdens^\top) = 
\nonumber\\
=(\mathbf{I}-\mathbf{1}\vecdens^\top) + 
\sum_{k=1}^\infty (\matprobtrans-\mathbf{1}\vecdens^\top)^k\,,
\nonumber
\end{eqnarray}
where $\matprobtrans^\infty\equiv \lim_{k\rightarrow\infty}\matprobtrans^k$. Denoting with $\mathbf{A} = \matprobtrans-\mathbf{1}\vecdens^\top$ and knowing that $\matprobtrans^\infty=\mathbf{1}\vecdens^\top$ the sum 
\begin{equation}
\lim_{k\rightarrow\infty}\sum_{i=1}^{k} \mathbf{A}^{i} = \lim_{k\rightarrow\infty}\mathbf{A}(\mathbf{I}-\mathbf{A})^{-1}(\mathbf{I}-\mathbf{A}^{k+1}) = \mathbf{A}(\mathbf{I}-\mathbf{A})^{-1}\,,
\nonumber
\end{equation}
since from Eq.~(\ref{eq:power_identity}) $\lim_{k\rightarrow\infty}\mathbf{A}^k=\mathbf{0}$. 

Finally, we have
\begin{equation}
\mathbf{B} = (\mathbf{I}-\mathbf{1}\vecdens^\top) + 
(\matprobtrans-\mathbf{1}\vecdens^\top)(\mathbf{I}-\matprobtrans+\mathbf{1}\vecdens^\top)^{-1}\,,
\end{equation}
and hence the infinite sum of auto-covariances from Eq.~(\ref{eq:cov}) is given by 
\begin{align}
\sum_{\tau=1}^\infty \mathrm{Cov}(l_1,l_{1+\tau}) &= 
\sum_{\tau=1}^\infty \vecdens^\top \mathbf{L}_1\, (\matprobtrans^{\tau-1}-\matprobtrans^{\infty})\mathbf{L}_1\,\mathbf{1} = \nonumber\\
&=\vecdens^\top\,\mathbf{L}_1\,\mathbf{B}\,\mathbf{L}_1\,\mathbf{1}\,.
\end{align}

Bringing everything together, the variance of individual edge lengths is
\begin{equation}
\sigma_l^2 = \langle l^2 \rangle - \langle l\rangle^2\ = 
 \vecdens^\top \mathbf{L}_2\, \mathbf{1} - (\vecdens^\top \mathbf{L}_1\, \mathbf{1})^2\,,
\end{equation}
where $(\mathbf{L}_2)_{ij}=\probtrans_{ij}\ln^2\probtrans_{ij}$,
the entropy rate can be expressed by using Eq.~(\ref{eq:l_av}) as 
\begin{equation}
S = \vecdens^\top \mathbf{L}_1 \mathbf{1}\,,
\label{eq:SMentropy_analytic}
\end{equation}
while the Lyapunov measure is given by the
\begin{equation}
\Lambda =  
   \vecdens^\top \mathbf{L}_2\, \mathbf{1} - (\vecdens^\top \mathbf{L}_1\, \mathbf{1})^2 + 2\vecdens^\top \mathbf{L}_1\, \mathbf{B}\, \mathbf{L}_1\, \mathbf{1}
\label{eq:SMlyapunov_analytic}
\end{equation}
closed-form expression, where $\vecdens^\top$ and $\mathbf{1} = (1, 1, \dots, 1)$ are the left and right eigenvectors of the stochastic transition matrix $\matprobtrans$ with eigenvalue $\alpha_1 = 1$, and
\begin{equation}
\mathbf{B} \equiv (\mathbf{I}-\matprobtrans+\mathbf{1}\vecdens^\top)^{-1} - \mathbf{1}\vecdens^\top\,.
\end{equation}
For systems defined by a random transition matrix $\matprobtrans$ with uniformly distributed elements, the Lyapunov-measure $\Lambda$ approaches 1/4 as the number of states goes to infinity~\cite{sandor2021}.

\subsection{Numerical results for prototypical examples}

\begin{figure}[ht!]
\includegraphics[width=0.8\columnwidth,keepaspectratio]{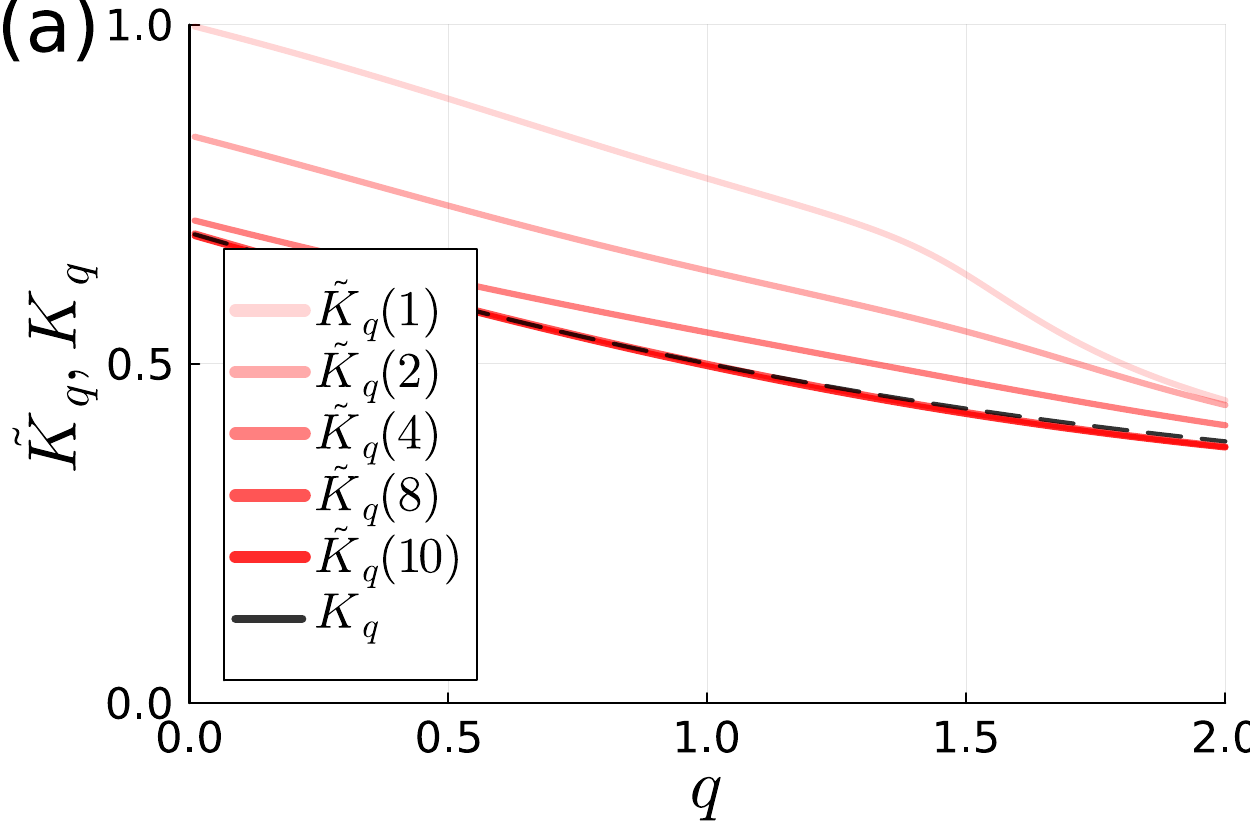}
\includegraphics[width=0.8\columnwidth,keepaspectratio]{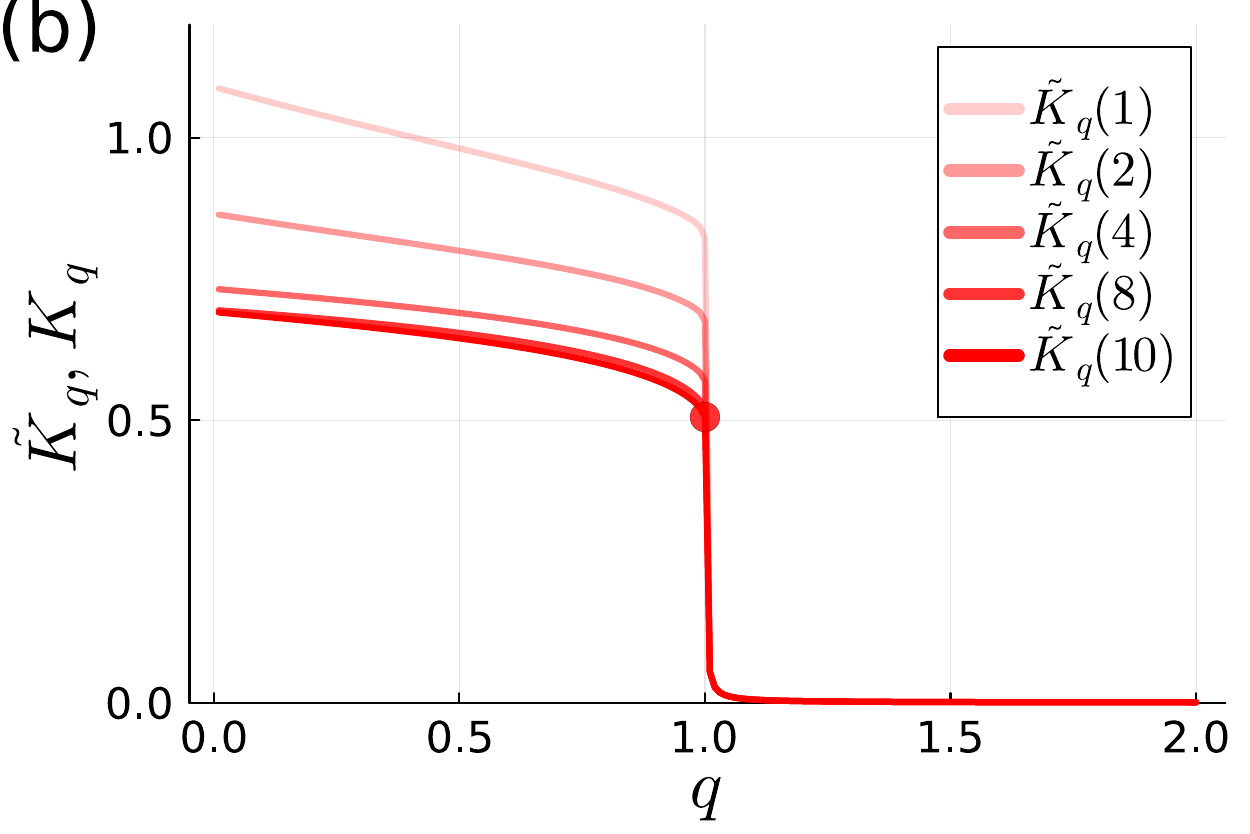}
\caption{\label{fig:renyi_entropy_spectrum} Spectrum of the truncated R\'enyi entropy  $\tilde{K}_q(m)$ of two different maps computed by using Eq.~(\ref{eq:renyi_entropy_estimation_derivation}) and states of order $m$ for trajectories of $10^8$ steps on the attractor, after discarding the first $10^6$ steps. 
(a) The asymmetric tent map defined by Eq.~(\ref{eq:tent}) with parameter $r=0.8$ using uniform partitioning to $n=2^{5}$ cells. The corresponding analytical entropy values, $K_q$, of Eq.~(\ref{eq:entropy_tent}) are shown by the dashed line. (b) The critical map from Eq.~(\ref{eq:critical}) with parameter $r=2$ using uniform partitioning to $n=2^{10}$ cells. The red marker denotes the $\tilde{K}_1(10)\approx K_1 = 0.5$ value \cite{szepfalusy1987}. 
For (a) and (b), the [0,1] interval has been partitioned into the given number of cells $n$.
}
\end{figure}

In addition to the examples in the main paper we consider the asymmetric tent map~\cite{beck1995thermodynamics}:
\begin{equation}
x_{\tau+1} = 
\begin{cases}
\dfrac{x_\tau}{r}\ ,& \quad\text{for } 0\leq x <r\,, \\
\\
\dfrac{1-x_\tau}{1-r}\ ,& \quad\text{for } r\leq x \leq 1\,,
\label{eq:tent}    
\end{cases}    
\end{equation}
where $r\in (0,1)$ and for which the $[0,r)$ and $[r,1)$ cells are generating and Markov partitions at the same time \cite{beck1995thermodynamics}. 
For this system, the analytical expression for the R\'enyi entropy is~\cite{beck1995thermodynamics}:
\begin{equation}
K_q = \ln(r^q+(1-r)^q)/(1-q)\ .
\label{eq:entropy_tent}
\end{equation}
Estimating the transition probability matrix $\matprobtrans$ from a sufficiently long trajectory of the system from Eq.~(\ref{eq:tent}) and using the Markov partitions from above, the entropies from Eqs.~(\ref{eq:entropy_tent}) and (\ref{eq:renyi_entropy_estimation_derivation}) coincide, $\tilde{K}_q=K_q$ (not shown here). For a partitioning composed of e.g.\ $n=2^5$ equally-sized cells combined with higher-order states, the truncated entropy in Eq.~(\ref{eq:renyi_entropy_estimation_derivation}) together with order $m=10$ provides a reasonably good estimate of the exact R\'enyi entropy spectrum given by Eq.~(\ref{eq:entropy_tent}), as shown in panel (a) of Fig.~\ref{fig:renyi_entropy_spectrum}.

Another example is the family of critical systems \cite{szepfalusy1987}:
\begin{equation}
x_{\tau+1} = 1-\left|x_\tau^r-(1-x_\tau)^r\right|^{1/r}\,,
\label{eq:critical}
\end{equation}
exhibiting intermittent dynamics. 
These systems present a dynamical phase transition at $q=1$~\cite{szepfalusy1987}. 
Constructing the Markov chain representation of the dynamics using $n=2^{10}$ equal-sized bins, and higher-order states with $m=10$, the R\'enyi entropy spectrum is computed using Eq.~(\ref{eq:renyi_entropy_estimation_derivation}) and shown in Fig.~\ref{fig:renyi_entropy_spectrum}(a). 
Though the absolute value of the truncated entropy $\tilde{K}_q$ depends on the resolution $n$ and the order $m$, the qualitative nature of the phase transition is unaffected (see Fig.~\ref{fig:renyi_entropy_spectrum}(b)). The exact value of the KS/metric entropy $S=0.5$~\cite{szepfalusy1987} is approximated well already for $m=10$ by $\tilde{K}_1$.

\begin{figure*}[hbt!]
\includegraphics[width=2\columnwidth,keepaspectratio]{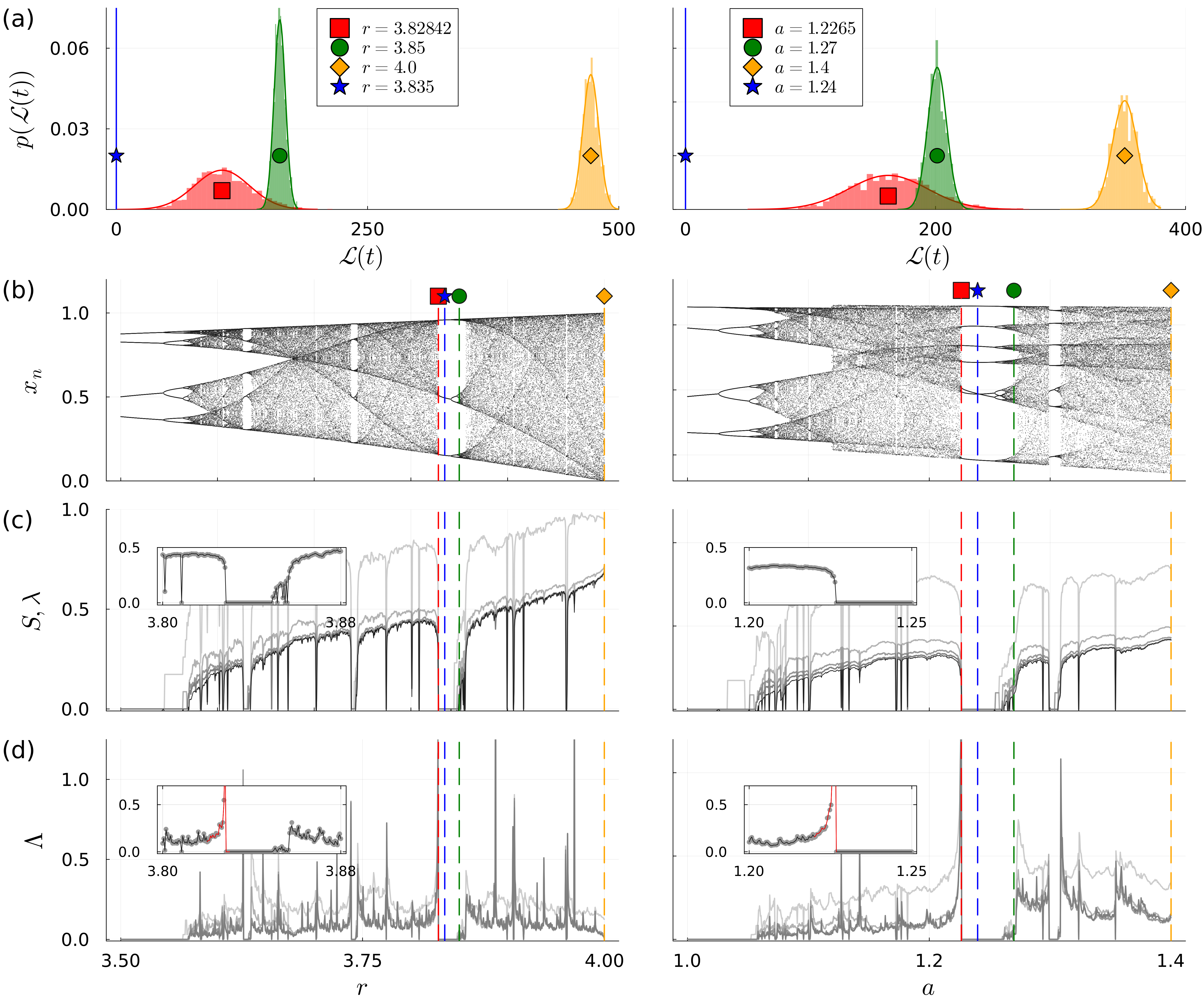}
\caption{\label{fig:henon_measures}
Matching the classical picture of the logistic map (left column) and the Henon map (right column) to its Markov-chain-based model. 
The first $10^6$ time steps preceding the $10^8$ total number of steps were discarded in order to eliminate the transients. 
For the logistic map, the [0,1] interval is partitioned into $n=2^5$ equally sized cells and iterated for 1000 different values of the control parameter $r$ covering the [3.5, 4] range. 
Similarly, for the Henon map the phase plane $[-2, 2]\times[-2, 2]$ is partitioned into a $n = 2^5 \times 2^5$ grid and iterated for 1000 different values of the control parameter $a$ covering the [1.0, 1.4] range.
The parameter values $r$ and $a$ used as examples in panels (a) and (b) of Fig. 1 of the main paper, respectively, are indicated by the corresponding markers and the vertical dashed lines.
(a) Demonstration of the effect of the central-limit theorem for Markov chains on the total path lengths, $\mathcal{L}(t)$, in the limit $t\rightarrow\infty$. 
Histograms are obtained from 10$^3$ random walk simulations of length $t=500$ on the corresponding $m=1$ (first-order) state-transition networks (STN).
Continuous lines represent the normal distributions with mean $\langle\mathcal{L}(t)\rangle = S(m=1)\cdot t$ and variance $\Delta^2\mathcal{L}(t) = \Lambda(m=1)\cdot t$, where the KS entropy $S$ and Lyapunov measure $\Lambda$ are computed based on Eqs.~(\ref{eq:SMentropy_analytic}) and (\ref{eq:SMlyapunov_analytic}). 
(b) The bifurcation (orbit) diagrams of the two systems.  
(c) The KS entropy, $S(m)$, and (d) the Lyapunov measure, $\Lambda(m)$. Lines in increasingly darker shades of gray correspond to Markov chains using states of order $m = 1,2,4,12$, respectively. 
In panel (c) the largest Lyapunov exponent, $\lambda$, is also shown by the thin black line for comparison.
Note that $S(12)\approx\lambda$ for $\lambda>0$.
The analytical results (gray lines) given in Eq.~(\ref{eq:SMlyapunov_analytic}) are compared to the statistics of 1000 simulated random walks of 500 steps (gray circular markers) on the corresponding STNs in the insets for order $m=12$ with the divergent $\Lambda$ values preceding the bifurcations marked by the red curves.
}
\end{figure*}

Fig.~\ref{fig:henon_measures} provides further details on the logistic and Henon maps compared to Fig. 1 in the main paper. 
Panel (a) demonstrates numerically the effect of central-limit theorem for Markov chains on the total path lengths, $\mathcal{L}(t)$, in the limit of $t\rightarrow\infty$. 

\subsection{Numerical estimation of $S$ and $\Lambda$}

The computational complexity of evaluating $S$ in Eq.~(\ref{eq:SMentropy_analytic}) and the first two terms of $\Lambda$ in Eq.~(\ref{eq:SMlyapunov_analytic}) are of $\mathcal{O}(N^2)$ where $N$ is the number of symbols. 
The third term contains a matrix inverse but in conjunction with the inner product allowing for the application of a linear solver instead of inversion. 
More precisely we are looking for the vector 
\[
\mathbf{z} = (\mathbf{I}-\matprobtrans+\mathbf{1}\vecdens^\top)^{-1}\mathbf{L}_1 \mathbf{1}\;.
\] 
Since the number of symbols increases exponentially with the order of the representation the computational cost of the inverse calculation becomes prohibitive for high orders.  
However, in this case, $\matprobtrans$ becomes a sparse matrix facilitating the use of high-efficiency iterative solvers. 
The solution in $\mathbf{z}$ is obtained as the fix-point of the iteration
\[
\mathbf{z}' = \matprobtrans\mathbf{z} -\mathbf{1}\langle z\rangle +\mathbf{L}_1 \mathbf{1}\;.
\]
where $\langle z\rangle\equiv \vecdens^\top\mathbf{z}$.
Using a random vector as an initial value of $\mathbf{z}$ the iteration is convergent for all systems studied here. 


%


\subsection{CODE AVAILABILITY}
The full code base has been released on Github, see Ref.~\cite{github}. The basic functionalities as computing the symbolic representation, the transition matrix, and the corresponding measures are implemented in our Julia package, StateTransitionNetworks.jl. Main results included in this work can be repeated from scratch with the codes in the 'stn\_paper' directory (see the attached README file).


\bibliography{stn_paper}